\newcommand{\beq}{\begin{equation}}
\newcommand{\eeq}{\end{equation}}
\newcommand{\beeq}{\begin{eqnarray}}
\newcommand{\eeeq}{\end{eqnarray}}

\documentstyle[12pt,epsfig] {article}

\begin{document}

\begin{center}\bf
CONFORMAL  COVARIANT DESCRIPTION  OF MESONS AND BARYONS  IN COMPOSITE SUPERCONFORMAL STRING MODEL  

\bigskip
V. A.  Kudryavtsev  \\  
    Petersburg  Nuclear Physics Institute\\
\bigskip
A b s t r a c t
\end{center}

   Operator nucleon vertices are constructed in composite superconformal string model.
Splitting of baryon Regge trajectories with the same quantum numbers but with  opposite parity is provided by 
inclusion of simple additional component to superconformal generators. 

\section{Introduction. }

            String models  (previously dual resonance models) \cite{1,2} had appeared as 
 a possible way  to describe  strong  interactions at low and intermediate energies.
Presently phenomelogical status of such appproach seems to be even more impressive  than fourty years ago
since we have now stringlike spectrum of hadron states including not only leading Regge trajectories but  and
second and third daughter Regge trajectories for this spectrum \cite{3,4} up to spins equal to be
to 2,3 and even up to 4 for mesons or 11/2 for baryons .  However before  we  had not
consistent realistic string amplitudes for hadrons without negative norm states in physical spectrum and with
 intercept of leading meson ($\rho$) trajectory to be equal to one half \cite{5}.  It turns out to be possible 
to build  a composite string model to be compatible with these requirements \cite{6}. This model  gives 
 realisic description of meson spectrum and brings to  correct interaction of $\pi$ -mesons  which satisfies
the Adler-Weinberg condition for soft $\pi$ -mesons \cite{7}.   

  This composite string construction  arises from some
 generalization of classical multi-reggeon (multi-string) vertices \cite{8}.
  New generalized multi-reggeon  vertices
 were  suggested  in 1993   as a new solution of duality equations  in \cite{8}.
Corresponding  string amplitudes have been used for
description of many $\pi$-mesons interaction \cite {9}.
This approach gives  a new geometric picture for string interaction. It
has a natural description in terms of  three two-dimensional
surfaces (composite string) instead ofl
one two-dimensional surface in classical case \cite{2}.
We have new additional two-dimensional fields on these additional edging two-dimensional surfaces for this composite string .
   Zeroth components of these new two-dimensional fields
have as eigenvectors  colourless spinors $\lambda^{(i)}$  which
carry quark quantum numbers (momentum, flavor, spin) for this $(i)$ edging surface. This composite
string construction reminds two other similar objects: a gluon
string with two point-like quarks at its ends or simplest case of a
string ending at two branes when they are themselves some strings.
Let us note that we have no supersymmetry in the target space (Minkowski space) for this model.
 There is the conformal supersymmetry
(super Virasoro symmetry) on two-dimensional world surfaces  only. The
topology of interacting composite strings allows to solve the
problem of the intercept $\alpha (0)$ to be equal to one half for the leading meson Regge trajectory.
 It gives consistent  description of string amplitudes which do not break down
unitarity at tree level  if  the super Virasoro (superconformal) symmetry and some additional symmetry to be generated by supercurrent conditions act (see \cite {10}). 

\section{General Formulation of Composite Superconformal String Model.}

Many-string vertices of interacting composite strings generalize  naturally
 many-string (many-reggeon) vertices for classical string models \cite{8}.
 A $N$-string amplitude is represented by some integral over $z_i$ variables of the vacuum
expectation value for a product of a vertex operator $V_N (a_n^{(i)})$ and wave
functions of string states $\hat \Psi^{(i)}(a_n^{(i)+})$:
\begin{eqnarray}
A_N= \int \prod_j dz_j \langle 0|V_N \prod_i \hat \Psi^{(i)}|0^{(i)}\rangle\,
\end{eqnarray}
where $\hat \Psi^{(i)}(a_n^{(i)+})|0^{(i)}\rangle\ $ is a wave function of $i$-th string state;
$[a_{n\mu}^{(i)},a_{m\nu}^{(j)+}]=g_{\mu\nu}\delta_{mn}\delta_{ij}$.

The {\it N}-string vertex operator $V_N^{NS}$ for the classical Neveu--Schwarz model is given by the following exponent:
\begin{eqnarray}
 V_N^{NS}=\exp \frac{1}{2}\sum_{n,m,p; i\ne k}\frac{a_n^{(i)}}{\sqrt n}
(U_{\varepsilon}^{(i)})_{nm}(V_{\varepsilon}^{(k)})_{mp}\frac{a_p^{(k)}}{\sqrt p}\ +\\ \nonumber
+\frac{1}{2} \sum_{n,m,p; i\ne k}
b^{(i)}_{n+1/2}(U_{1/2}^{(i)})_{nm}(V_{1/2}^{(k)})_{mp}b^{(k)}_{n+1/2}\, 
\end{eqnarray}
 where $\{b_{r\mu}^{(i)},b_{s\nu}^{(j)+}\}=g_{\mu\nu}\delta_{rs}\delta_{ij}$;
$(U_{j}^{(i)})_{nm}$, $(V_{j}^{(k)})_{mp}$ are the
definite infinite matrices. The value of index $j$ corresponds to
representation of these matrices according to certain conformal spin
$j$ of two-dimensional fields: $j=\varepsilon\!\to\!0$ for the first item and
$j=\frac{1}{2}$ for the second one.\cite{11} The upper indices
$i,k=1...N$ number interacting strings. These vertices $V_N^{NS}$
have the necessary factorization and conformal properties and
satisfy duality property .

It turns out there is another  operator $W_N$ with  these properties and  to be satisfying duality equations:
\begin{eqnarray}
W_N=\sum_{n,m,k} \tilde \Psi^{(1)}_{n+\frac{1}{2}}(U_{1/2}^{(1)})_{nm}(V_{1/2}^{(2)})_{mk}\Psi^{(2)}_{k+\frac{1}{2}} \sum_{l,p,s}\tilde \Psi^{(2)}_{l+\frac{1}{2}}(U_{1/2}^{(2)})_{lp}(V_{1/2}^{(3)})_{ps}\Psi^{(3)}_{s+\frac{1}{2}} \times \dots \cr \dots \times \sum_{q,r,f}\Psi^{(N)}_{q+
\frac{1}{2}}(U_{1/2}^{(N)})_{qr}(V_{1/2}^{(1)})_{rf}\Psi^{(1)}_{f+\frac{1}{2}} \ \\ \nonumber
W_N = \prod_{i=1}^N \sum_{n,m,p}\tilde \Psi^{(i)}_{n+\frac{1}{2}}(U_{1/2}^{(i)})_{nm}(V_{1/2}^{(i+1)})_{mp}\Psi^{(i+1)}_{p+\frac{1}{2}}
\end{eqnarray}
This operator $W_N$ is a cyclic symmetrical trace-like operator built of Fourier components of two-dimensional fields $\Psi ^{(i)}$.
 The generalized {\it N}-string vertex operator for composite strings is the product of the old operator $V_N^{NS}$ and the new one $W_N$:
\begin{eqnarray}
V^{comp}_N=W_NV_N^{NS}.
\end{eqnarray}
It is obvious that operators $V_N^{NS}$ and $W_N$ have different structure. 
Matrices $U^{(i)}$ and $V^{(j)}$ in $V_N^{NS}$ connect all possible $i$-th fields components $a_n^{(i)}$ with each other $j$-th fields components.
  However, in the operator $W_N$ matrices $U^{(i)}$ and $V^{(i+1)}$ connect only neighbor fields $\Psi^{(i)}$ and $\Psi^{(i+1)}$ .

That is why this operator reproduces the structure of dual quark diagrams. These diagrams display the well known OZI (Okubo--Zweig--Iizuka) rule. 
So operator (4) can be interpreted as vertex operator of composite strings interaction.
 It is worth to  be  stressed that all many-string vertex operators (2),(3),(4) possess necessary cyclic symmetry and duality properties.  Duality properties  automatically lead
 to  satisfaction of one-particle unitarity in {\it s}- and {\it t}-channel, if physical states spectrum is free of negative norm states.

For investigation of composite superconformal string it is more
convenient to move from multi-string vertices to the emission of the ground state. Then the
composite superconformal string vertex operator will contain
additional (to usual $d X_{\mu }$ and $H_{\mu }$ fields on the basic
two-dimensional surface) fields on the edging surfaces: $Y_{\mu }$ and
its superpartner $f_{\mu }$ with Lorentz indices $\mu =0,1,2,3$. We also
include other fields (corresponding to internal quantum numbers) {\it
J} and its superpartner $\Phi$ on edging surfaces and similar {\it I} and $\theta $ fields on the basic
two-dimensional surface.

 Since the edging fields are propagating only on the own edging surface between neighboring vertexes
 we will have to label these fields in accordance with numbers of these edging surfaces. 
This is similar to numbering of strings in many-string vertexes (\cite{8}). For the ground state emission  
 the vertex operator $\hat V_{i,i+1}$ will contain the fields of $i$-th and $(i\!+\!1)$-th edging surfaces.  
The following description of ground states interaction amplitudes will be:
\begin{eqnarray}
A_N=\int\prod dz_i\langle0|\hat V_1(z_1)\hat V_2(z_2)\dots\hat V_{N-1}(z_{N-1})\hat V_N(z_N)|0\rangle, \cr \hat V_i(z_i)=z_i^{-L_0}V_i(1)z_i^{L_0}, \cr \hat V_i(1)=g(\tilde\Psi_i(1)\Gamma_i\Psi_{i+1}(1)e^{ik_iX(1)}),
\end{eqnarray}
where $\Psi_i$ is the field propagating on the $i$-th edging surface.
We formulate here the vertex $\hat V_{i,i+1}$ corresponding to the emission of ground state in an amplitude $A_N$ for {\it N} strings interaction. The vertex operator $\hat V_{i,i+1}$ for $\pi$-meson emission has the simplest form for this composite string model:
\begin{eqnarray}
\hat V_{i,i+1}(z_i)=z_i^{-L_0} \left[ G_r,\hat W_{i,i+1} \right] z_i^{L_0}, \\
\hat W_{i,i+1} = \hat R_i^{out}\hat R_{NS}\hat R_{i+1}^{in}=\\ \nonumber
=:e^{(ik_{i }Y^{(i)}(1)+i\xi_{i}J^{(i)}(1))}\tilde\lambda ^{(i)+}_{\alpha}\Gamma^{i,i+1}_{\alpha\beta}\lambda ^{(i+1)-}_{\beta}\\ \nonumber e^{(ik_{i,i+1}X(1)+i\zeta_{i,i+1} I(1))}e^{(ik_{i+1}Y^{(i+1)}(1)+i\xi_{i+1}J^{(i+1)}(1))}: .
\end{eqnarray}

        Here for pi-meson vertex  $ \tilde\lambda ^{(i)+}_{\alpha}\Gamma^{i,i+1}_{\alpha\beta}\lambda ^{(i+1)-}_{\beta}=
\tilde\lambda ^{(i)+}\gamma_5 \tau^{(i,i+1)}\lambda ^{(i+1)-}$.
The operators $\hat R_i^{out}$ and $R_{i+1}^{in}$ are defined by fields on $i$-th and $(i\!+\!1)$-th edging surfaces. 
The operator $\hat R_{NS}$ is defined by fields on the basic surface. They have the same structure as the operator $W_{NS}$ of classical Neveu--Schwarz string model:
\begin{eqnarray}
W_{NS}=:\exp ip_iX(1):=\exp(-p_i\sum_n\frac{a_{-n}}{n})\exp(-ip_iX_0)\exp(p_i\sum_n\frac{a_{n}}{n})
\end{eqnarray}
     Namely  we have for  R operators the following expressions:
\begin{eqnarray}
\hat R_i^{out}=\exp(\xi_i\sum_n\frac{J^{(i)}_{-n}}{n})\exp(k_i\sum_n\frac{Y^{(i)}_{-n}}{n})\exp(ik_i\bar Y_0^{(i)})
\tilde\lambda ^{(i)+}_{\alpha} \times \cr \times\exp(-k_i\sum_n\frac{Y^{(i)}_{n}}{n})\exp(-\xi_i\sum_n\frac{J^{(i)}_{n}}{n}),
\end{eqnarray}
\begin{eqnarray}
R_{i+1}^{in}=\exp(-\xi_{i+1}\sum_n\frac{J^{(i+1)}_{-n}}{n})\exp(-k_{i+1}\sum_n\frac{Y^{(i+1)}_{-n}}{n})\exp(-ik_{i+1}
\bar Y_0^{(i+1)})\lambda ^{(i+1)-}_{\beta}\times \cr \times \exp(k_{i+1}\sum_n\frac{Y^{(i+1)}_{n}}{n})\exp(\xi_{i+1}\sum_n\frac{J^{(i+1)}_{n}}{n}),
\end{eqnarray}

\begin{eqnarray}
\hat R_{NS}=\exp(-\zeta_{i,i+1}\sum_n\frac{I_{-n}}{n})\exp(-p_{i,i+1}\sum_n\frac{a_{-n}}{n})\exp(-ip_{i,i+1}X_0)\times \cr \times\Gamma^{i,i+1}_{\alpha\beta}\exp(p_{i,i+1}\sum_n\frac{a_{n}}{n})\exp(\zeta_{i,i+1}\sum_n\frac{I_{n}}{n}).
\end{eqnarray}

Here we have introduced $\lambda_{\alpha}$ operators to carry quark flavors and quark spin degrees of freedom. It must be stressed that $\lambda^{(i)}$ is an eigenvector of the operator $J_0^{(i)}$ and therefore it is an analog of $\exp(ik_iX_0)$ for field $d X$ and $\exp(ik_i\bar Y_0^{(i)})$ for field $Y^{(i)}$, since $\langle \lambda^{(i)}| J_0^{(i)}=\xi_i\langle \lambda^{(i)}| $ as $(\exp(ik_iX_0))\hat P=k_i(\exp(ik_iX_0))$ and $(\exp(ik_i\bar Y_0^{(i)}))Y_0^{(i)}=k_i(\exp(ik_i\bar Y_0^{(i)}))$. This approach replaces usual transition to extra dimensions and allows introduce the quark quantum numbers in a natural way. In addition,  we obtain an attractive interpretation of the Chan-Paton factor in terms of two-dimensional fields.
\begin{eqnarray}
\langle0|\tilde\lambda^{(+)}=0,\\ \lambda^{(-)}|0\rangle=0.
\end{eqnarray}
\begin{eqnarray}
\left\{ \tilde\lambda^{(-)}_{\alpha},\lambda^{(+)}_{\beta}\right\}=\delta_{\alpha,\beta},\\\tilde\lambda=\lambda T_0,\\T_0=\gamma_0\otimes\tau_2.
\end{eqnarray}
                             
     From (9),(10) and (11) we see that the expression $\tilde\lambda ^{(i)+}_{\alpha}\Gamma^{i,i+1}_{\alpha\beta}\lambda ^{(i+1)-}_{\beta}$
with  $ \Gamma^{i,i+1}= \gamma_5 \tau^{i,i+1}$   plays a role of $\pi$-meson function to be composed as a quark - antiquark  pair $\tilde\lambda ^{(i)+} \lambda ^{(i+1)-}$.

       Also were used values:
\begin{eqnarray}
p_{i,i+1}=k_{i+1}-k_{i}, \\ \zeta_{i,i+1}=\xi_{i+1}-\xi_{i};. 
\end{eqnarray}

Here we have to fulfill conditions for momenta: ${k_i^2=k_{i+1}^2=0}$, $k_ik_{i+1}=0$. These conditions lead to $m_{\pi}^2=0$ 
and necessary supercurrent conditions (see\cite{10}).

 We require $\xi_i^2=\xi_{i+1}^2=1/2$ in order to have the conformal spin of this vertex to be equal to one and necessary superconformal
symmetry.

Using this vertex operator ((6),(7), (9-11)) we can write $A_{\pi\pi}$ as the following integral:
\begin{eqnarray}
 A_{\pi\pi}=
-g^2Tr(\Gamma_{12}\Gamma_{23}\Gamma_{34}\Gamma_{41})\int_0^1 dz\;
z^{-1}z^{\frac{1}{2}(-p_{13}^2 + \xi_{1}^2+\xi_{3}^2-k_{1}^2-k_{3}^2+\zeta_{13}^2)}\times
\cr \times
(1-z)^{-p_{23}p_{34}+\zeta_{23}\zeta_{34}+k_{3}^2-\xi_{3}^2-1}(-p_{23}p_{34}+\zeta_{23}\zeta_{34}+k_{3}^2-\xi_{3}^2).
\end{eqnarray}
So we obtain this amplitude as a simple beta function:
\begin{eqnarray}
 A_4 = - g^2 \frac{\Gamma (1- \alpha^t_0-\frac1{2}t)\Gamma
(1- \alpha^s_0-\frac1{2}s)}
{\Gamma(1-\alpha^t_0-\frac1{2}t-\alpha^s_0-\frac1{2}s))}
Tr(\Gamma_{12}\Gamma_{23}\Gamma_{34}\Gamma_{41})
\end{eqnarray}
  with $t=p_{13}^2$, $s=p_{34}^2$. Now we take into account $k_i^2=k_{i+1}^2=0$, $\xi_i^2=\xi_{i+1}^2=1/2$ and $\zeta_{13}=\zeta_{23}=\zeta_{34}=\xi_{3}-\xi_{4}=0$;
$p_{23}^2=p_{34}^2=m_{\pi}^2=0$ and derive $\alpha^t_0=\alpha^s_0=\frac1{2}$. So finally we have:
\begin{eqnarray}
A_{\pi\pi}=-g^2Tr(\Gamma_{12}\Gamma_{23}\Gamma_{34}\Gamma_{41})\frac{\Gamma(1-\alpha_t^{\rho})\Gamma(1-\alpha_s^{\rho})}{\Gamma(1-\alpha_t^{\rho}-\alpha_s^{\rho})}
\end{eqnarray}
with $\alpha_t^{\rho}=\frac{1}{2}+\frac{t}{2}$ and $\alpha_s^{\rho}=\frac{1}{2}+\frac{s}{2}$.

        It is worth  to note that
we can give another equivalent form of vertex operator $\hat V_i(1)$  in (5,6,7)
 without enumeration for $\Psi$  fields. It is $ Y^{(i)},J^{(i)}$ fields above.  For one set of $\Psi$ fields it brings to the following expression:
\begin{eqnarray}
\hat V^{i,i+1}(1)=g(\tilde\Psi(1)\Gamma_{(i,i+1)}|0^{\Psi(1)}\rangle\
\langle0^{\Psi}|\Psi(1)e^{(ik_{i,i+1}X(1)+i\zeta_{i,i+1} I(1))},
\end{eqnarray} 
The operator $|0^{\Psi}\rangle\ \langle\ 0^{\Psi}|$ allows fields $\Psi$
to be propagating between neighboring vertices only. Due to these
operators the form (6) displays  factorization in terms of finite numbers of fields and excludes composite string model from
the set of classical additive string models of the Lovelace's paper \cite{5}.
This construction reveals  the topology of composite string model.

\section {Treatment of nucleons in composite superconformal string model and elimination of degeneracy in mass of parity doublets for baryons.}

          Inclusion of baryons in composite superconformal string  model requires new operator vertices and correspoding amplitudes. 

           For baryon we  
consider  two-dimensional fields on the basic surface and on three additional two-dimensional surfaces in accordance with  a new expression
 $\lambda ^{(i)} \lambda ^{(f)}\lambda ^{(i+1)}$ for a baryon wave function instead of two additional surrfaces and the form  $\tilde\lambda ^{(i)} \lambda ^{(i+1)}$ for meson case
 in the  preceding section. Third (f) baryon  surface  goes between   the (i)-th  and (i+1)-th edging surfaces.
Again this topology repeats the topology of dual quark diagrams for the case of baryons.

 As it is well-known usual fermion poles are determined by the propagators:
\begin{equation}
 \frac {\hat{q}+m}{q^2-m^2} \qquad \mbox{for} \qquad \frac{1}{2}^{+}
\end{equation}
\begin{equation}
 \frac {-\hat{q}+m}{q^2-m^2}= \gamma_5\frac {\hat{q}+m}{q^2-m^2}\gamma_5 \qquad \mbox{for} \qquad \frac{1}{2}^{-}
\end{equation}
Here
\begin{equation}
 \frac {\hat{q}+m}{2m}= \Pi_{+},\qquad
  \frac {-\hat{q}+m}{2m}= \Pi_{-}
\end{equation}
$\Pi_{+}$; $\Pi_{-}$ are the projectors for parity $P=\!+1$ and $P=\!-1$

For arbitrary $q^2=u$ (off mass shell) we have

\begin{equation}
 \frac {\hat{q}+q}{2q}= \Pi_{+} \qquad \mbox{and} \qquad \frac {-\hat{q}+q}{2q}= \Pi_{-}
\end{equation}
here  $q=\sqrt{q^2}$

          The Gribov-MacDowell symmetry  for Regge fermion trajectories with
opposite parities \cite{12}:

\begin{equation}
    \alpha^{+}(q)=\alpha^{-}(-q)
\end{equation}
and for Regge pole vertices:
\begin{equation}
\Gamma^{+}(q)=\Gamma^{-}(-q)
\end{equation}

 removes  a singularity
 in the scattering amplitudes  for $q^2=0$.

             For classical string amplitudes usually we have linear in $q^2$
trajectories:  ${\alpha(q^2)=\alpha_{0}+\alpha'q^2}$ without the
$\hat{q}$ dependence.  And hence we obtain the degeneracy in mass
for fermion states of opposite parities with obvious contradiction
with most experimental data for the baryon spectrum.
 Superstring and Ramond approaches \cite {2} do not allow to solve this problem
 of the baryon spectrum description in
the context of a string treatment and they are
 incompatible with the observed spectrum of  baryons.

        It is proved that introduction of a new two-dimensional field with zeroth component to be proportional  $ I^{P}_0 = A +B\overline{\lambda}^{(+)}_{f}\hat{P}\lambda^{(-)}_{f}$ as in
recent papers  (see  \cite{13})  can not  solve this problem  in the framework of the
composite superconformal string model since it breaks down conformal and superconformal properties of corresponding vertices and amplitudes.
But it is possible  more gentle and superconformally covariant solution  of the elimination of degeneracy   in mass for parity twins wthout new two-dimensional field with above mentioned charge $ I^{P}_0$ and infinite number of other components.

      Namely we add some simple operators $\Delta$ to usual superconformal generators $G_r$ to be determining constraints for physical states.  We determine this procedure  in  the following way:

\begin{equation}
 \tilde G_r |Phys.state\rangle=0; r>0 \\ \nonumber
\end{equation}
\begin{equation}
\tilde G_r= G_r+ \Delta_r ;
\end{equation}
 r is a half-integer number;
\begin{equation}
\Delta_r=  (\overline{\lambda}^{(+)}_{f}b(\hat{P}-m_N)\lambda^{(-)}_{f}) \delta _r\\  \nonumber
\end{equation}
\begin{equation}
\delta_r =\frac 1{\xi_{(f)}}\Phi_r^{(f)} - \frac 1 {m_N^2}(k_{(f)}  f_r^{(f)})
\end{equation}
 
    Here    $\Phi_r^{(f)}$  and  $ f_r^{(f)}$ are r-th components   of  anticommuting two-dimensional  fields $ \Phi^{(f)}$ and  $ f_{\mu}^{(f)}$  on  fermion additional surface.
\begin{equation}
\{ \Delta_r, G_s\}=-(r+s)(\overline{\lambda}^{(+)}_{f}b(\hat{P}-m_N)\lambda^{(-)}_{f})(\frac 1{\xi_{(f)}}J_{r+s}^{(f)} - \frac 1 {m_N^2}(k_{(f)} Y_{r+s}^{(f)}))
\end{equation}
      Here    $J_n^{(f)}$  and  $Y_n^{(f)}$ are n-th components   of commuting two-dimensional  fields $J^{(f)}$ and $ Y_{\mu}^{(f)}$ to be superpartners of  $ \Phi^{(f)}$ and  $ f_{\mu}^{(f)}$- fields.   

       And  new vertices  $V(z)$ are defined as:
\begin{equation}
 V(z_i)=z_i^{-\tilde L_0} \left[\tilde G_r, W(1) \right] z_i^{\tilde L_0} \\
\end{equation}

    here
\begin{equation}
\tilde L_0=\frac{1}{2}\{\tilde G_{\frac{1}{2} } , \tilde G_{-\frac{1}{2} }\}= L_0 +\Delta_0; \quad  \tilde L_0 |Phys.state\rangle= \frac 1{2}; 
\end{equation}
\begin{equation}
 \Delta_0 =\{\Delta_r ,  \Delta_{-r}\}=(\overline{\lambda}^{(+)}_{f}B(\hat{P}-m_N)\lambda^{(-)}_{f})^2
\end{equation}

 Meson vertices in composite superconformal model are constructed as superconformal  vertices of Neveu-Schwarz
 (NS)-type \cite{14} for pi -mesons 
and Bardakci-Halpern (BH)-type \cite{15} for  K-meson vertices.
Quark spinors are included in vertices as eigenvectors for charges (
zeroth components) of two-dimensional fields on the corresponding
edging surfaces.

(NS)-type vertices consist odd number of anticommutating components,
they are $V_{-}$ vertices (G-parity odd). (BH)-type vertices consist even number of
anticommutating components, they are  $V_{+}$ vertices (G-parity even). 
The $\pi$ meson vertices are (NS)-type vertices  . Hence the nucleon
vertex has to have both types vertex operators
  in order to obtain  nonvanishing transitions $N\overline{N}$ to both even number of pions 
(G-parity even) and to odd number of pions (G-parity odd)  but  it is enough to include in
nucleon vertex  $(a+c\delta )$ factor which mixes different G-parity  contributions   due to
the anticommuting component $ \delta $. It is worth  to note that vertices of pi-meson emission from fermions will contain  $\delta $ component according  to the equation (34).

So we have
\begin{equation}
V^{(N)}=V^{+} + V^{-}.
\end{equation}

and  the elimination  of parity doublets degeneracy in the baryon spectrum
 since we have   here  the superconformal operator $L_0(\hat{P})$ and 
the corresponding mass condition $L_0+\Delta_0 =\frac 1{2}$ to be
dependent on $\hat{P}$.

          In subsequent papers we consider properties of amplitudes for  interactions of pions, K-mesons  and nucleons to be based on these vertices.

      I would like to thank  my colleague A.N.Semenova
 for useful discussions of this work.


\begin{thebibliography} {99} \fussy
\bibitem{1} Y.Nambu, Lectures at the Copenhagen  Symposium "Symmetries and
quark models" (1970);\\
 G.Veneziano, Nuovo cim.{\bf 57A} (1968) 190;
\bibitem{2} M.B.Green,J.H.Schwarz and E.Witten, Superstring theory (Cambridge Univ.Press, N.Y.1987);\\
\bibitem{3} C.Lovelace, Phys Lett. {\bf B28} (1968) 264;
            D.V.Shirkov, Sov.Phys.Usp. {\bf 102} (1970)87;\\
\bibitem{4} A.V.Anisovich, V.V.Anisovich and A.V.Sarantsev, Phys.Rev.{\bf D62} (2000) 051502;
            V.V.Anisovich, Phys.Usp. {\bf 47} (2004) 45; \\
\bibitem{5} C.Lovelace, Nucl.Phys. {\bf B148} (1979) 253;\\
\bibitem{6} V.A.Kudryavtsev, JETP Lett.{\bf 58} (1993) 321; \\
 \bibitem{7} S.L.Adler, Phys Rev. {\bf B137} (1965) 1022;\\           
\bibitem{8}V.Alessandrini, D.Amati, M.Le Bellac and D.Olive,
 Phys.Rep. {\bf1} (1971) 269;\\
\bibitem{9}V.A.~Kudryavtsev,
 Phys.At.Nucl.{\bf 58} (1995) 131;\\
\bibitem{10} V.A.Kudryavtsev,  Nucl.Phys.Proc.Suppl. {\bf B198} (2010) 228;hep-th/1003.3943; \\
                   V.A.Kudryavtsev, A.N.Semenova, IJMPA. {\bf 27} (2012) 1250170228;\\
\bibitem{11}
E.Corrigan, C.Montonen,
 Nucl.Phys. {\bf B36} (1972) 58;
\bibitem{12}
S.W.MacDowell,
 Phys.Rev. {\bf 116} (1959) 774.
\bibitem{13}
 V.A.Kudryavtsev, A.N.Semenova,
Theoretical and Mathematical Physics,
{\bf 176} (2013) 922;\\
\bibitem{14}
A.Neveu and J.H.Schwarz,
 Nucl.Phys. {\bf B31} (1971) 86;
\bibitem{15}
K.Bardakci and M.B.Halpern,
 Phys.Rev. {\bf D3} (1971) 2493.

\end{thebibliography}
\end{document}